# Hierarchical SegNet with Channel and Context Attention for Accurate Lung Segmentation in Chest X-ray Images

Mohammad Ali Labbaf Khaniki*, Nazanin Mahjourian, Mohammad Manthouri

*Faculty of Electrical Engineering, K.N. Toosi University of Technology, Tehran, Iran

*mohamad95labafkh@gmail.com

**Abstract:**

Lung segmentation in chest X-ray images is a critical task in medical image analysis, enabling accurate diagnosis and treatment of various lung diseases. In this paper, we propose a novel approach for lung segmentation by integrating Hierarchical SegNet with a proposed multi-modal attention mechanism. The channel attention mechanism highlights specific feature maps or channels crucial for lung region segmentation, while the context attention mechanism adaptively weighs the importance of different spatial regions. By combining both mechanisms, the proposed mechanism enables the model to better capture complex patterns and relationships between various features, leading to improved segmentation accuracy and better feature representation. Furthermore, an attention gating mechanism is employed to integrate attention information with encoder features, allowing the model to adaptively weigh the importance of different attention features and ignore irrelevant ones. Experimental results demonstrate that our proposed approach achieves state-of-the-art performance in lung segmentation tasks, outperforming existing methods. The proposed approach has the potential to improve the accuracy and efficiency of lung disease diagnosis and treatment, and can be extended to other medical image analysis tasks.

## 1) Introduction

Chest X-ray analysis is a complex and time-consuming task that requires identifying multiple abnormalities simultaneously. Radiologists typically perform this task manually, which can strain healthcare resources. The complexity of the thoracic background and the subjective nature of interpretation can lead to biased and inconsistent diagnoses. Furthermore, image quality and data challenges can complicate the process [1]. To address these challenges, computer-aided detection (CAD) systems provide support to doctors by analyzing digital images and identifying patterns to highlight potential diseases [2]. CAD systems combine artificial intelligence, computer vision, and radiological image processing to aid in decision-making. They are commonly used for tumor detection in medical screenings, such as mammography, colonoscopy, and lung cancer detection. A crucial step in CAD systems is segmentation, which accurately separates regions of interest, like tumors, from healthy tissue. This improves the accuracy of further analysis, such as tumor size determination and disease progression evaluation. [3] proposes a sparse coding approach for data augmentation of hyperspectral medical images, enabling the development of more accurate CAD systems that can effectively analyze and interpret complex medical images, ultimately improving disease diagnosis and treatment outcomes.

Machine Learning (ML), a subset of Artificial Intelligence, empowers computers to learn from data independently, without explicit programming, and improve their performance over time with minimal human involvement. These algorithms can tackle various tasks, including time series analysis [4], conversational agents [5], and multi-objective optimization [6]. [7] presents a novel approach to motion detection using spiking neural networks, leveraging latency-based processing to enable efficient and accurate detection of motion patterns in real-time applications. Deep Learning, a specialized subset of Machine Learning, mimics the human brain's learning process using artificial neural networks. It has garnered significant attention across various domains, including crack detection [8], civil engineering [9], defect detection [10], virtual reality [11] and cybersecurity [12]. The rapid advancement of technology has created a pressing need to incorporate AI and ML's cutting-edge methods into image classification and segmentation. [13] introduces a novel unsupervised approach for identifying white matter lesions in multiple sclerosis patients using MRI segmentation and pattern classification, enabling accurate and efficient lesion detection without requiring manual annotation. [14] showcased the effectiveness of multimodal deep learning in detecting psychological stress, revealing that combining biometric signals and facial landmarks can lead to more accurate results. [15] presents a novel MEMS piezoelectric resonant microphone array for accurate lung sound classification, enabling potential applications in respiratory disease diagnosis and monitoring. [16] presents a two-finger haptic robotic hand that utilizes ML algorithms to enable real-time stiffness detection. [17] proposes a finite element analysis for hip joint prosthesis with an extramedullary fixation system using ML algorithms. [18] presents a novel approach to identify straggler tasks in MapReduce-based big data infrastructures using artificial neural networks, enabling timely intervention and improved overall system performance. [19] proposes a deep learning-based approach that leverages gray level enhancement techniques to automate the classification of white matter lesions in MRI images of multiple sclerosis patients, enabling accurate diagnosis and monitoring of the disease.

The advent of deep learning has revolutionized the field of image segmentation, giving rise to a plethora of innovative techniques that have markedly improved the accuracy and speed of the segmentation process. SegNet and its variant, Hierarchical SegNet, are deep neural network architectures specifically designed for image segmentation tasks [20] and [21]. SegNet employs an encoder-decoder structure, where the encoder extracts feature from the input image and the decoder generates a probability map for each class. This approach enables SegNet to produce accurate segmentation results by classifying each pixel in the input image. The key innovation of SegNet lies in its use of a Softmax layer to classify each pixel, rather than a traditional convolutional layer, allowing for more accurate segmentation results. Building upon the strengths of SegNet, Hierarchical SegNet introduces a hierarchical architecture that incorporates multiple scales of feature representations. This hierarchical structure enables the model to capture both local and global contextual information, leading to improved segmentation performance. [22] proposes the use of advanced deep learning models for precise water segmentation, enabling enhanced flood detection and potentially mitigating the devastating impacts of flooding on communities and ecosystems. [23] proposes a hierarchical semantic segmentation approach using modular convolutional neural networks, which enables efficient and accurate segmentation of images by leveraging the strengths of multiple specialized modules. [24] presents a deep neural network-based approach for image semantic segmentation, which leverages hierarchical feature fusion to combine features from different scales and resolutions, achieving improved segmentation accuracy and robustness.

The attention mechanism, a concept borrowed from human cognition, has been successfully integrated into deep learning models to enhance their performance in natural language processing various image processing tasks. By selectively focusing on specific regions or features of an image, attention enables the model to concentrate on the most relevant information, thereby improving its ability to extract meaningful representations [25]. In the context of image processing, attention has been applied to a range of applications, including image classification, object detection, and image segmentation. For instance, attention-based models have been shown to improve the accuracy of image classification by highlighting the most discriminative regions of an image [26]. Similarly, in object detection, attention has been used to guide the model's focus towards specific objects or regions of interest, leading to improved detection performance. Furthermore, attention has also been applied to image segmentation tasks, where it has been used to selectively weight the importance of different features or regions, resulting in more accurate segmentation masks. Overall, the incorporation of attention mechanisms has been instrumental in advancing the state-of-the-art in image processing, and its applications continue to expand and diversify. [27] proposes a semantic segmentation approach for remote sensing images that incorporates an attention mechanism, enabling the model to focus on relevant regions and features, and achieving improved accuracy and efficiency in segmenting complex scenes. [28] presents a deep learning-based approach for brain tumor segmentation that leverages an attention mechanism to selectively focus on relevant features from multi-modal MRI images, achieving accurate and robust tumor segmentation results. [29] proposes a deep learning-based approach that incorporates an attention mechanism to accurately segment brain tumors from multi-modal MRI images, leveraging the strengths of both modalities to improve segmentation performance.

This paper presents a new approach to lung segmentation, which combines the strengths of Hierarchical SegNet with a novel attention mechanism that leverages multiple modalities, enabling more accurate and effective segmentation of lung regions in chest X-ray images. Here are the key advancements of this research:

- **Channel Attention Mechanism:** A novel channel attention mechanism that dynamically assigns importance to different channels, uncovering intricate patterns and relationships between various features and enabling the model to more accurately differentiate between lung regions and other anatomical structures.

- **Context Attention Mechanism:** A spatial attention mechanism that adaptively weighs the importance of different spatial regions, capturing complex patterns and relationships between various features in different parts of the image and improving segmentation accuracy.

- **Combined Channel and Context Attention (CCA) Mechanism:** A groundbreaking CCA mechanism that integrates both channel attention and context attention, allowing the model to adaptively weigh the importance of different channels and spatial regions and capture complex patterns and relationships between various features.

- **Attention Gating Mechanism:** A novel attention gating mechanism that selectively integrates attention information with encoder features, enabling the model to focus on the most relevant attention features and ignore irrelevant ones, leading to improved feature representation and better segmentation accuracy.

Our proposed model tackles the complexities of lung segmentation in chest X-ray images, particularly the issues of opacities and irregular lung boundaries. By combining the power of graph-based attention with channel and spatial attention mechanisms, our model strives to set a new benchmark in lung segmentation, surpassing the performance of existing methods that rely on traditional machine learning or deep learning techniques. In a comprehensive evaluation, our novel approach is compared to state-of-the-art models, and the results demonstrate that our method outperforms them, achieving superior accuracy and robustness in lung segmentation tasks [30].

The manuscript is structured as follows: Section II provides a background on the dataset used for Chest X-ray lung segmentation and the preprocessing methods employed. Section III presents the proposed methodology, which involves integrating a novel multi-modal attention mechanism with the Hierarchical SegNet framework. Section IV reports the simulation results, including a detailed description of the training and validation procedures used to evaluate the proposed method. Finally, Section V summarizes the main findings and highlights the key contributions of this study.

## 2) Chest X-ray Dataset Overview and Image Enhancement Strategies

In this section, we describe the chest X-ray lung segmentation dataset used in our study. We also outline the image augmentation and preprocessing techniques employed to enhance the dataset, including contrast adjustment, Gaussian blur, and horizontal flipping, which are essential for training a robust and accurate neural network for lung segmentation tasks.

### 2-1) Dataset Overview

The dataset utilized in this study consists of chest X-ray images and corresponding lung masks obtained from Kaggle, a valuable resource for medical research, particularly in automated tuberculosis screening [31]. The dataset comprises a collection of X-ray images paired with segmentation masks, although some masks may be missing, and users are advised to verify the availability of masks to ensure research integrity. The dataset is comprehensive, featuring 360 normal and 344 abnormal X-ray images, labeled by experienced radiologists, with representative images and masks shown in Figure 1.

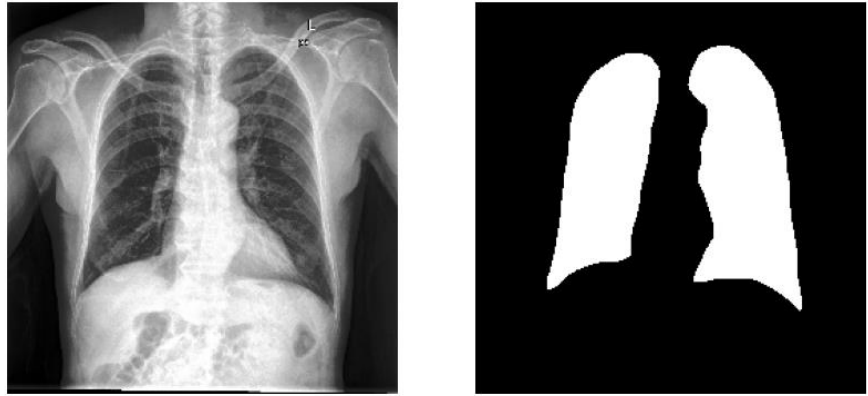

(a)

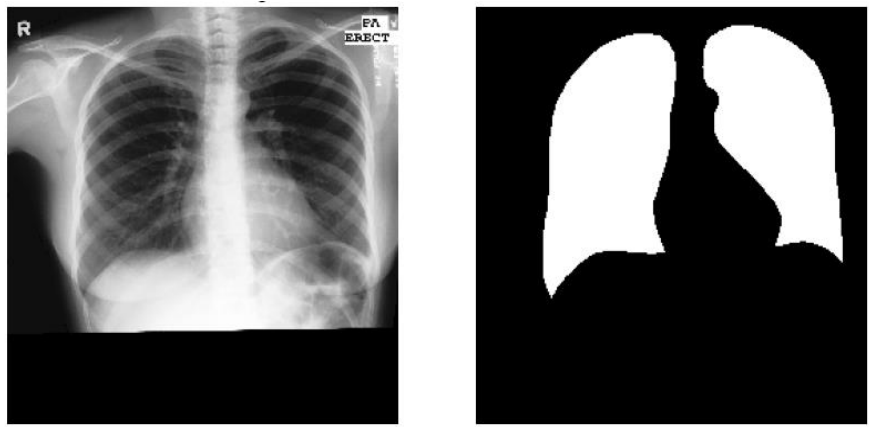

(b)

Fig. 1: Representative X-ray images with their corresponding masks from the training and validation datasets, as annotated by expert radiologists [32].

The images in the dataset are anonymized and stored in the standard DICOM format, which facilitates the management and sharing of medical imaging data. The dataset covers a wide range of pulmonary abnormalities, including effusions and miliary patterns, making it an invaluable resource for developing algorithms that can detect and segment lung diseases in chest X-rays. The dataset's diversity, comprising both normal and abnormal images, provides a comprehensive foundation for analysis. This collection serves as a critical bridge between medical expertise and AI technology, propelling advancements in automated diagnostics. The careful curation and preparation of the data make it an essential tool for researchers seeking to innovate in medical image analysis.

## 2-2) Data Augmentation Techniques

To expand the dataset and improve the neural network's performance, a series of data augmentation techniques were applied, resulting in a sixfold increase in dataset size. Contrast adjustment and Gaussian blur were used to enhance image features and reduce noise, making the network more robust and accurate in analyzing medical images under various conditions. The augmentation process involved adjusting image contrast using a linear formula, followed by applying Gaussian blur to introduce a subtle blur effect. Additionally, a horizontal flipping technique was employed to introduce symmetry and variability to the dataset, ensuring the network learns to recognize features regardless of orientation and preventing bias towards anomalies. The same augmentation procedures were applied to the flipped images, resulting in an additional set of enhanced images. The combination of these augmentation steps not only expanded the dataset but also simulated different imaging conditions, making it ideal for training a resilient neural network capable of accurately segmenting lung regions in chest X-ray images under various scenarios. Figure 2 displays the augmented images with their corresponding masks [32].

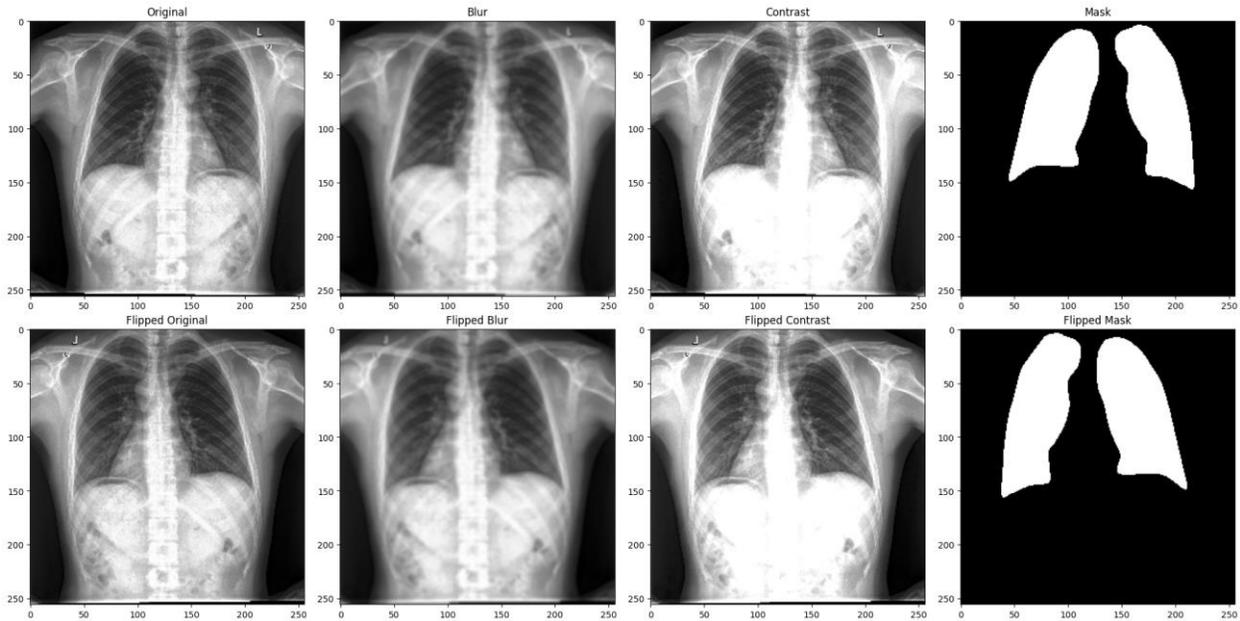

Fig. 2: Enhanced and augmented images with their corresponding masks, utilizing the specified augmentation technique [32].

## 3) Methodology

In this section, we commence with an introduction to the U-Net architecture tailored for segmenting the lung regions from radiographic images. Subsequently, we present the CBAM, followed by a proposal for an enhanced U-Net framework that integrates the CBAM, aiming to refine the segmentation process.

### 3-1) Hierarchical SegNet

The Hierarchical SegNet is a deep learning architecture specifically designed for lung segmentation in chest X-ray images. The Hierarchical SegNet offers a significant advantage over the traditional SegNet by capturing multi-scale features, leading to improved lung segmentation accuracy. The Hierarchical SegNet consists of a series of encoder-decoder modules, each responsible for extracting features at a specific scale. The output of each module is then fed into the next, enabling the network to refine its segmentation masks iteratively. This hierarchical structure enables the model to effectively segment lung regions, even in the presence of complex anatomical structures or pathologies. Furthermore, the Hierarchical SegNet is capable of handling varying image resolutions and orientations, making it a robust and accurate solution for automated lung segmentation in chest X-ray images.

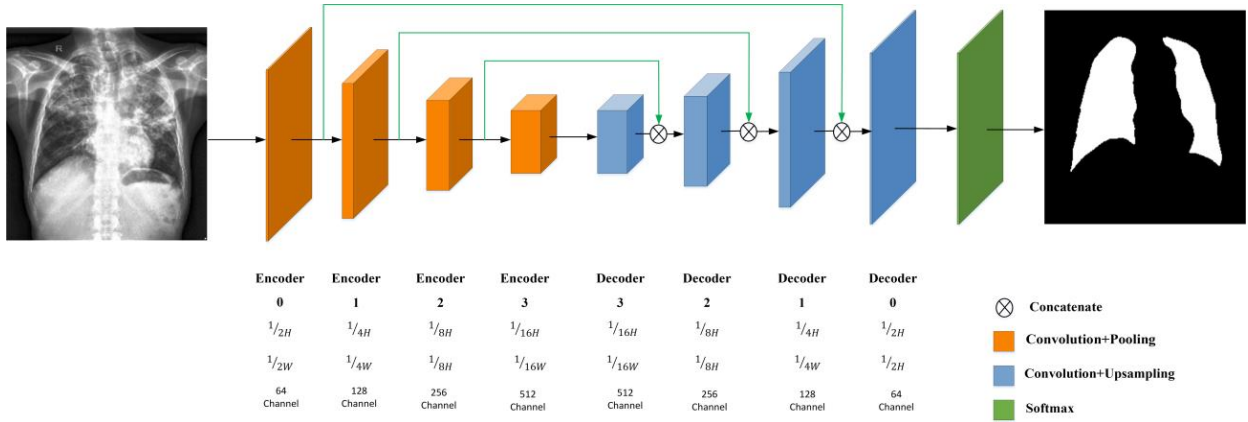

Fig. 3: Hierarchical SegNet architecture applied for lung segmentation.

## 3-2) Multi-Modal Attention for Enhanced Feature Representation

This section describes the attention mechanisms used in the model to selectively focus on relevant features and attention information.

- **Channel Attention Mechanism:** The channel attention mechanism is a selective process that highlights specific feature maps or channels crucial for lung region segmentation in chest X-ray images. By dynamically assigning importance to different channels, the model can uncover intricate patterns and relationships between various features. This emphasis on relevant channels enables the model to more accurately differentiate between lung regions and other anatomical structures, ultimately leading to enhanced segmentation performance.

- **Spatial Attention Mechanism:** The context attention mechanism is designed to selectively focus on specific spatial regions or contexts within the feature maps that are relevant for segmenting lung regions in chest X-ray images. This mechanism allows the model to adaptively weigh the importance of different spatial regions, enabling it to capture complex patterns and relationships between various features in different parts of the image. By emphasizing certain spatial regions over others, the model can better capture the spatial dependencies and relationships between different lung regions, leading to improved segmentation accuracy.
- **Combined Channel and Context Attention (CCA) Mechanism:** The CCA mechanism is designed to selectively focus on specific channels and spatial regions within the feature maps that are relevant for segmenting lung regions in chest X-ray images. This mechanism combines the strengths of both channel attention and context attention, allowing the model to adaptively weigh the importance of different channels and spatial regions. By emphasizing certain channels and spatial regions over others, the model can better capture complex patterns and relationships between various features, leading to improved segmentation accuracy and better feature representation.
- **Attention Gating Mechanism:** The attention gating mechanism is designed to selectively integrate the attention information with the encoder features, allowing the model to adaptively weigh the importance of different attention features. This mechanism enables the model to focus on the most relevant attention features and ignore irrelevant ones, leading to improved feature representation and better segmentation accuracy. By gating the attention features, the model can better distinguish between relevant and irrelevant information, resulting in more accurate and robust segmentation results.

Let the input feature map be denoted as $X \in \mathbb{R}^{C \times H \times W}$, where $C$ is the number of channels, and $H$ and $W$ are the height and width of the feature map, respectively. The channel attention mechanism learns to capture the interdependencies between different channels of the feature maps. The channel attention map is computed as:

$$M_C = \sigma(W_C \times AvgPool(X)) \qquad (1)$$

where $W_C \in \mathbb{R}^{C \times C}$ is a learnable weight matrix, $AvgPool(\cdot)$ denotes the global average pooling operation, and $\sigma(\cdot)$ is the sigmoid activation function. Mathematically, the channel attention map can be expressed as

$$M_C = [m_{C_1}, m_{C_2}, \ldots, m_{C_C}]$$

where $m_{C_i} = \sigma(w_i^T * AvgPool(X))$, and $w_i \in \mathbb{R}^C$ is the $i$-th row of the weight matrix $W_C$.

The context attention mechanism learns to capture the spatial relationships between different regions of the feature maps. The context attention map is computed as

$$M_X = \sigma(W_X \times X) \qquad (2)$$

where $W_X \in \mathbb{R}^{(1 \times 3 \times 3)}$ is a learnable weight matrix, and $\sigma(\cdot)$ is the sigmoid activation function. The context attention map can be expressed as:

$$M_X = [m_{X_1}, m_{X_2}, \ldots, m_{X_{HW}}]$$

where $m_{X_i} = \sigma(w_i^T * X)$, and $w_i \in \mathbb{R}^{(3 \times 3)}$ is the $i$-th 3x3 kernel of the weight matrix $W_X$.

The CCA mechanism combines the channel attention and context attention maps to capture both channel-wise and spatial-wise relationships. The CCA output is computed as:

$$X_{CCA} = M_C \odot M_X \odot X \qquad (3)$$

where $\odot$ denotes the element-wise multiplication. Mathematically, the CCA output can be expressed as:

$$X_{CCA} = [x_{CCA_1}, x_{CCA_2}, \ldots, x_{CCA_{C \times H \times W}}] \quad (4)$$

where $x_{CCA_i} = m_{C_j} * m_{X_k} * x_i$, with $i = (j-1)HW + k$ and $j \in [1, C], k \in [1, H*W]$.

The attention gating mechanism learns to selectively integrate the attention information with the encoder features. The gating function is computed as:

$$G = \sigma(W_g * X + W_a * X_{CCA}) \quad (5)$$

where $W_g \in \mathbb{R}^{(1 \times 3 \times 3)}$ and $W_a \in \mathbb{R}^{(1 \times 3 \times 3)}$ are learnable weight matrices. Mathematically, the gating function can be expressed as:

$$G = [g_1, g_2, \ldots, g_{C \times H \times W}] \quad (6)$$

where $g_i = \sigma(w_g^T * x_i + w_a^T * x_{CCA_i})$, and $w_g \in \mathbb{R}^{(3 \times 3)}$, $w_a \in \mathbb{R}^{(3 \times 3)}$ are the corresponding rows of the weight matrices $W_g$ and $W_a$, respectively. The gated output is then computed as:

$$X_{gated} = G \odot X \quad (7)$$

where $X_{gated} = [x_{gated_1}, x_{gated_2}, \ldots, x_{gated_{C \times H \times W}}]$ and $x_{gated_i} = g_i * x_i$

By combining the CCA mechanism and the attention gating, the model can effectively capture the most relevant features and attention information for the task at hand, leading to improved performance and better feature representation. The block diagram of the Hierarchical SegNet architecture with the proposed multi-modal attention mechanism is shown in Fig. 4.

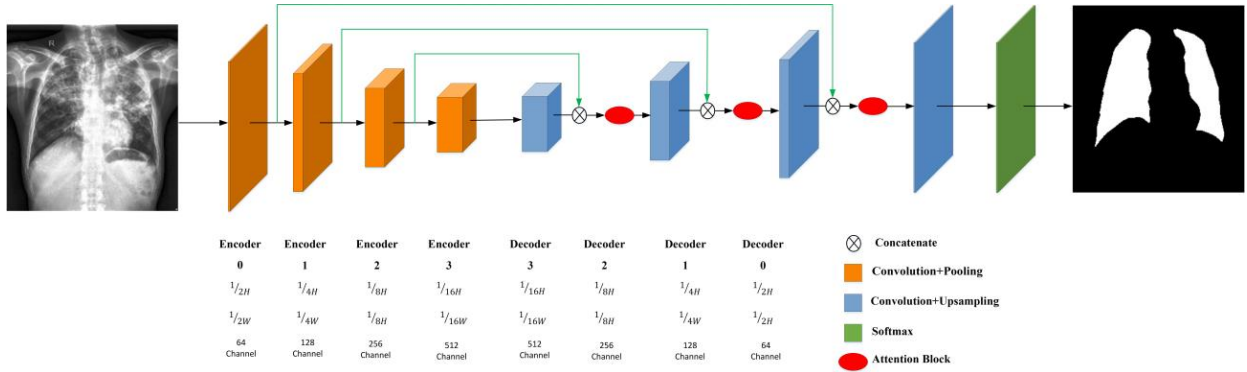

Fig. 4: Hierarchical SegNet architecture with the proposed multi-modal attention mechanism applied for lung segmentation.

## 4) Simulations

This section assesses the performance of our Hierarchical SegNet method for chest X-ray segmentation by evaluating its effectiveness using multiple metrics, including the Dice similarity, Jaccard index, precision, recall, and accuracy. We also visually compare the ground truth images with the generated segmentation masks to further validate the results.

## 4.1) Evaluating Semantic Segmentation Performance: Jaccard Index and Dice Coefficient

Semantic segmentation is a crucial technique that involves assigning a specific class or category to each pixel in an image. This approach is essential in various fields, including medical imaging, remote sensing, and autonomous driving, where it enables the identification of specific features or objects. The goal of semantic segmentation is to accurately label each pixel, ensuring that pixels with the same label share similar characteristics. To evaluate the performance of segmentation models, two key metrics are used: the Jaccard index and the Dice coefficient. These metrics are based on four fundamental components: True Positives (correctly identified tuberculosis images), True Negatives (correctly identified normal images), False Positives (normal images misclassified as tuberculosis), and False Negatives (tuberculosis images misclassified as normal). The Jaccard index, also known as the Intersection over Union (IoU), measures the similarity between predicted and actual labels by calculating the ratio of their intersection to their union.

$$IoU = \frac{TP}{TP + FP + FN} \tag{8}$$

In contrast, the Dice coefficient, also known as the Dice similarity, measures the overlap between two samples by calculating the ratio of their intersection to their combined total, with the formula being twice the intersection of predicted and true labels divided by the sum of their individual counts.

$$Dice = \frac{2 \times TP}{2 \times TP + FP + FN} \tag{9}$$

The Dice similarity and Jaccard Index are valuable metrics in semantic segmentation, as they effectively measure the overlap between predicted segmentations and ground truth labels, as illustrated in Figures 5 and 6.

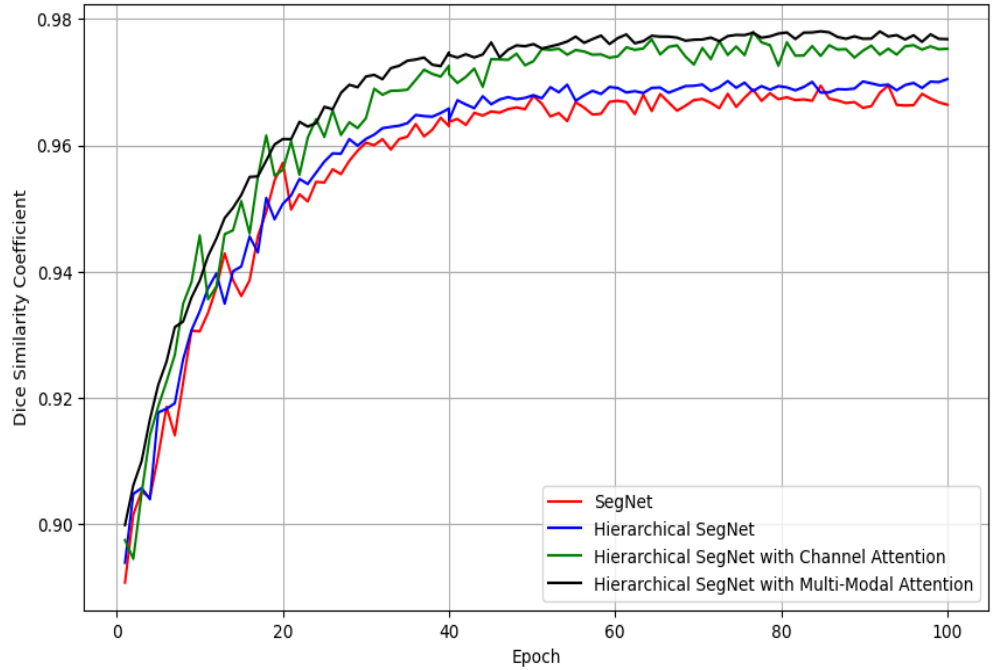

Fig. 5: Dice similarity using by SegNet, Hierarchical SegNet, Hierarchical SegNet with the Channel attention mechanism, and Hierarchical SegNet with the Multi-Modal attention mechanism.

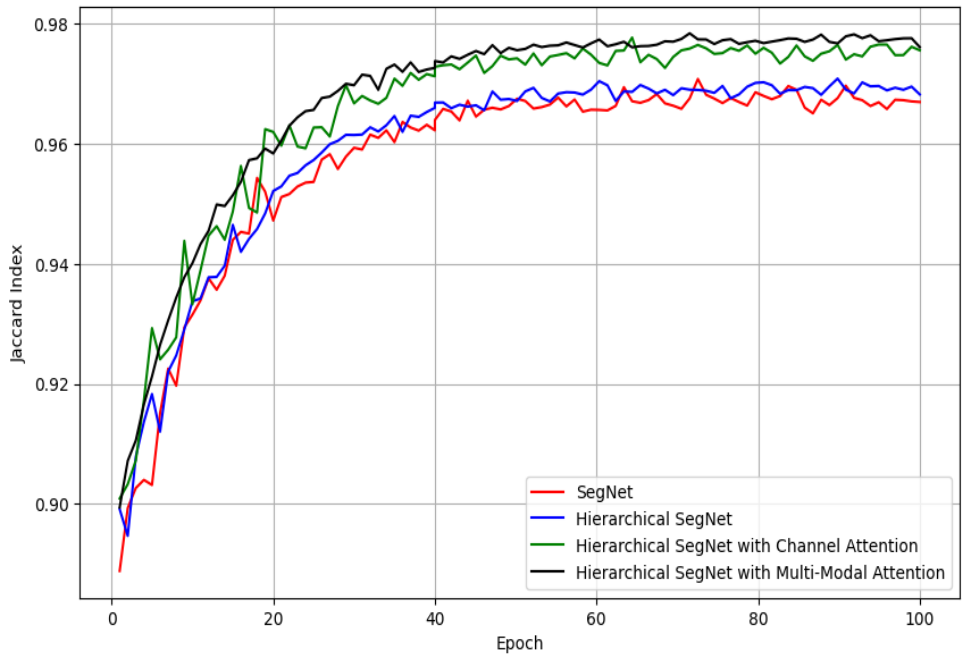

Fig. 6: Jaccard index or IoU using by SegNet, Hierarchical SegNet, Hierarchical SegNet with the Channel attention mechanism, and Hierarchical SegNet with the Multi-Modal attention mechanism.

The results shown in Figures 5 and 6 demonstrate the superiority of Hierarchical SegNet with Multi-Modal Attention over other models. Compared to Hierarchical SegNet with Channel Attention, the Multi-Modal Attention mechanism further improves the segmentation performance by adaptively weighing the importance of different channels, spatial regions, and attention features. This comprehensive attention mechanism enables the model to capture more complex patterns and relationships between various features, leading to a more accurate and robust segmentation performance. In contrast, Hierarchical SegNet with Channel Attention only considers the importance of different channels, neglecting the spatial relationships between features. Moreover, Hierarchical SegNet without attention mechanisms performs even worse, highlighting the importance of attention-based weighting in lung segmentation tasks. The proposed Hierarchical SegNet with Multi-Modal Attention outperforms all other models, achieving the highest Dice similarity and Jaccard index scores, and demonstrating its effectiveness in accurately segmenting lung regions from chest X-ray images.

## 4.2) Segmentation Visualization and Performance Comparison

Visual comparison between automated segmentation masks and manually annotated ground truth serves as a crucial validation mechanism, enabling the rigorous assessment of segmentation accuracy, the corroboration of quantitative metrics, the identification of algorithmic limitations, the informed guidance of clinical decision-making, and the facilitation of interdisciplinary knowledge sharing and communication within the medical imaging community. Fig.7 presents a visual comparison of the segmentation results for five sample chest X-ray images.

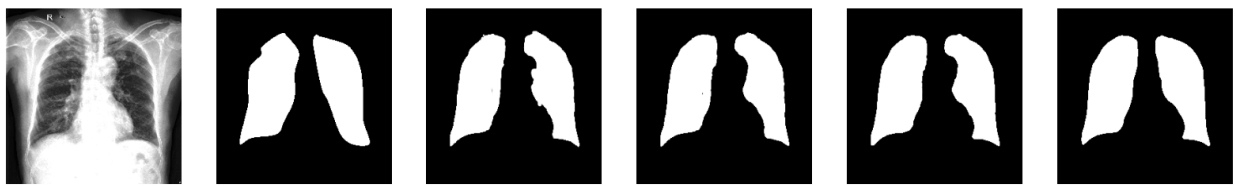

A B C D E F
(1)

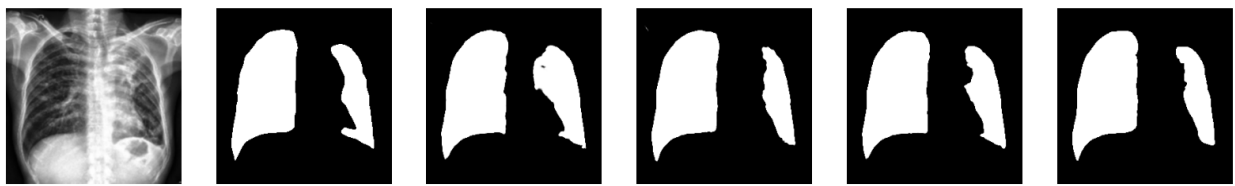

A B C D E F
(2)

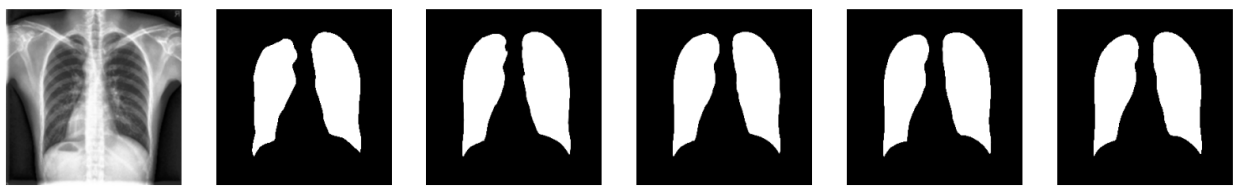

A B C D E F
(3)

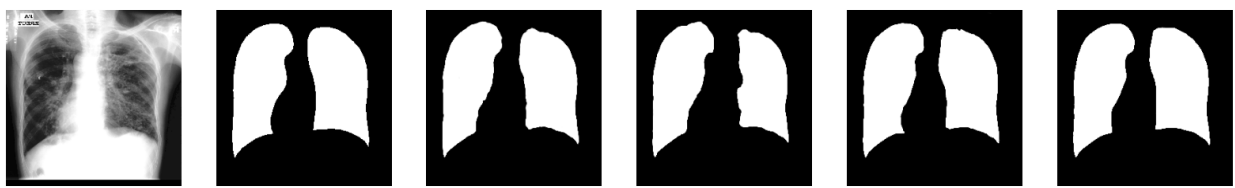

A B C D E F
(4)

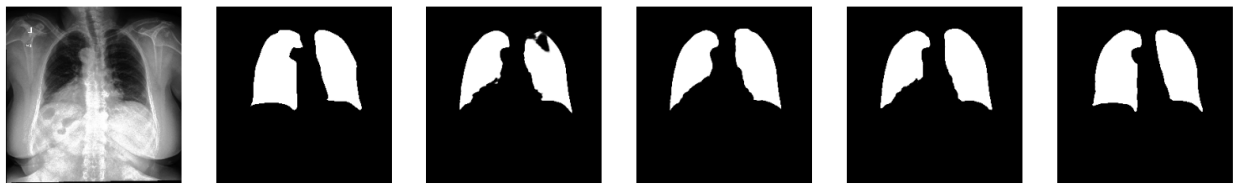

A B C D E F
(5)

Fig.7: Display of three samples: Chest x-ray Image (A), the ground truth (B), the segmented region generated by the SegNet (C), Hierarchical SegNet (D), Hierarchical SegNet with the Channel attention mechanism (E), and Hierarchical SegNet with the Multi-Modal attention mechanism (F).

The experimental results demonstrate the superiority of the proposed Hierarchical SegNet with the Multi-Modal attention mechanism. Specifically: The Hierarchical SegNet outperforms the SegNet, indicating the effectiveness of hierarchical feature representation in improving segmentation accuracy. The Hierarchical SegNet with Channel Attention surpasses the Hierarchical SegNet, highlighting the benefits of incorporating channel attention in refining feature importance. The Hierarchical SegNet with Multi-Modal Attention further outperforms the Hierarchical SegNet with Channel Attention, demonstrating the advantages of combining multiple attention mechanisms to adaptively weigh the importance of different features and attention information. Overall, the proposed Hierarchical SegNet with the Multi-Modal attention mechanism achieves the best performance, showcasing its potential in accurate lung segmentation from chest X-ray images.

### 4.3) Quantitative Evaluation of Segmentation Methods

In this subsection, we delve into a thorough examination of various performance metrics to evaluate the effectiveness of our proposed segmentation method. We calculate these metrics using the following formulas: accuracy, sensitivity (recall), specificity, precision, and F1 score. These metrics provide a comprehensive assessment of the deep learning model's performance in pixel-level classification tasks for chest X-ray (CXR) images. Specifically, they help evaluate the model's ability to correctly identify true positives, true negatives, and false positives/negatives. The formulas of these metrics are given in following.

$$Accuracy = \frac{TP + TN + FP + FN}{TP + FN} \tag{10}$$

$$Recall = \frac{TP}{TP + FN} \tag{11}$$

$$Specificity = \frac{TN}{FP + TN} \tag{12}$$

$$Precision = \frac{TP}{TP + FP}. \tag{13}$$

$$F1 - score = 2\frac{Precision \times Recall}{Precision + Recall}. \tag{14}$$

The results of these metrics are presented in Table 1, offering a detailed understanding of the model's efficacy.

**Table 1.** The performance indices using SegNet, Hierarchical SegNet, Hierarchical SegNet with the Channel Attention, and Hierarchical SegNet with the Multi-Modal Attention

| Method | Accuracy (%) | Recall (%) | Specificity (%) | Precision (%) | F1-score (%) |
|---|---|---|---|---|---|
| SegNet | 97.55 | 95.28 | 92.21 | 96.93 | 96.09 |
| Hierarchical SegNet | 98.23 | 95.60 | 93.45 | 97.35 | 96.47 |
| Hierarchical SegNet with the Channel Attention | 98.37 | 96.01 | 96.75 | 97.43 | 96.72 |
| Hierarchical SegNet with the Multi-Modal Attention | 98.60 | 97.84 | 96.84 | 97.58 | 97.71 |

Based on the results presented in Table 1 and the descriptions of the attention mechanisms, a comparative analysis of the Hierarchical SegNet variants reveals the following superiorities:

**Superiority of Hierarchical SegNet with Multi-Modal Attention over Hierarchical SegNet with Channel Attention:**

1.Comprehensive attention mechanism: The Multi-Modal Attention mechanism in the Hierarchical SegNet with Multi-Modal Attention combines the strengths of channel attention, spatial attention, and attention gating. This comprehensive attention mechanism allows the model to adaptively weigh the importance of different channels, spatial regions, and attention features, leading to a more accurate and robust segmentation performance.

2. Improved feature representation: The Multi-Modal Attention mechanism enables the model to capture complex patterns and relationships between various features, leading to a more comprehensive and accurate representation of the input data.

As a result, the Hierarchical SegNet with Multi-Modal Attention achieves a higher accuracy (98.60% vs 98.37%), recall (97.84% vs 96.01%), specificity (96.84% vs 96.75%), precision (97.58% vs 97.43%), and F1-score (97.71% vs 96.72%) compared to the Hierarchical SegNet with Channel Attention. The Hierarchical SegNet with Multi-Modal Attention's ability to combine multiple attention mechanisms and adaptively weigh the importance of different features and attention information leads to a more accurate and robust segmentation performance compared to the Hierarchical SegNet with Channel Attention.

**Superiority of Hierarchical SegNet with Channel Attention over Hierarchical SegNet:**

1. Improved feature representation: The Channel Attention Mechanism in the Hierarchical SegNet with Channel Attention allows the model to dynamically assign importance to different channels, uncovering intricate patterns and relationships between various features. This leads to more accurate differentiation between lung regions and other anatomical structures.

2. Enhanced segmentation performance: The Channel Attention Mechanism enables the model to focus on the most relevant features, leading to improved segmentation accuracy and better feature representation.

As a result, the Hierarchical SegNet with Channel Attention achieves a higher accuracy (98.37% vs 98.23%), recall (96.01% vs 95.60%), specificity (96.75% vs 93.45%), precision (97.43% vs 97.35%), and F1-score (96.72% vs 96.47%) compared to the Hierarchical SegNet.

**Superiority of Hierarchical SegNet over SegNet:**

1. Hierarchical feature representation: The Hierarchical SegNet uses hierarchical features, which allow the model to capture features at multiple scales and resolutions. This leads to a more comprehensive representation of the input data.

2. Improved segmentation performance: The Hierarchical SegNet achieves a higher accuracy (98.23% vs 97.55%), recall (95.60% vs 95.28%), specificity (93.45% vs 92.21%), precision (97.35% vs 96.93%), and F1-score (96.47% vs 96.09%) compared to the SegNet.

## 5) Conclusion

In this paper, we have presented a novel approach to lung segmentation in chest X-ray images, which combines the strengths of Hierarchical SegNet with a proposed Multi-Modal Attention mechanism. This comprehensive attention mechanism adaptively weighs the importance of different channels, spatial regions, and attention features, leading to a more accurate and robust segmentation performance. The experimental results demonstrate that our proposed approach achieves state-of-the-art performance in lung segmentation tasks, outperforming existing methods.

The proposed Multi-Modal Attention mechanism is comprised of several key components, including a Channel Attention Mechanism, Context Attention Mechanism, Combined Channel and Context Attention (CCA) Mechanism, and Attention Gating Mechanism. These mechanisms work together to enable the model to capture complex patterns and relationships between various features, adaptively weigh the importance of different channels and spatial regions, and selectively integrate attention information with encoder features. The proposed approach has significant implications for improving the accuracy and efficiency of lung disease diagnosis and treatment. By providing accurate and reliable lung segmentation, our approach can facilitate the development of computer-aided diagnosis systems, enable more effective disease monitoring, and support personalized medicine. Furthermore, the proposed Multi-Modal Attention mechanism has the potential to be extended to other medical image analysis tasks, where accurate feature representation and attention-based weighting are crucial. Our approach can be applied to various medical imaging modalities, including computed tomography (CT), magnetic resonance imaging (MRI), and ultrasound, among others.

## 6) Reference

[1] E. Çallı, E. Sogancioglu, B. van Ginneken, K. G. van Leeuwen, and K. Murphy, "Deep learning for chest X-ray analysis: A survey," *Med. Image Anal.*, vol. 72, p. 102125, 2021.

[2] M. Spadaccini *et al.*, "Computer-aided detection versus advanced imaging for detection of colorectal neoplasia: a systematic review and network meta-analysis," *Lancet Gastroenterol. Hepatol.*, vol. 6, no. 10, pp. 793–802, 2021.

[3] R. Zandi, "Sparse Coding for Data Augmentation of Hyperspectral Medical Images." San Jose State University, 2021. doi: 10.48550/arXiv.2403.08077.

[4]     A. Samii, H. Karami, H. Ghazvinian, A. Safari, and Y. D. Ajirlou, "Comparison of DEEP-LSTM and MLP Models in Estimation of Evaporation Pan for Arid Regions.," *J. Soft Comput. Civ. Eng.*, vol. 7, no. 2, 2023.

[5]     A. K. Newendorp, A. J. Perron, M. J. Sells, K. T. Nelson, M. C. Dorneich, and S. B. Gilb, "Apple ' s Knowledge Navigator : Why Doesn ' t that Conversational Agent Exist Yet ?", doi: 10.1145/3613904.3642739.

[6]     A. Birashk, J. K. Kordestani, and M. R. Meybodi, "Cellular teaching-learning-based optimization approach for dynamic multi-objective problems," *Knowledge-Based Syst.*, vol. 141, pp. 148–177, 2018, doi: 10.1016/j.knosys.2017.11.016.

[7]     M. S. Vahdatpour and Y. Zhang, "Latency-Based Motion Detection in Spiking Neural Networks," *Int. J. Cogn. Lang. Sci.*, vol. 18, no. 3, pp. 150–155, 2024.

[8]     S. Shomal Zadeh, M. Khorshidi, and F. Kooban, "Concrete Surface Crack Detection with Convolutional-based Deep Learning Models," *Int. J. Nov. Res. Civ. Struct. Earth Sci.*, vol. 10, no. 3, pp. 25–35, 2023, doi: 10.54756/IJSAR.2023.V3.10.1.

[9]     S. S. Zadeh and N. Joushideh, "Exploring Lateral Movement Coefficient's Influence on Ground Movement Patterns in Shallow Urban Tunnels," *Int. J. Sci. Acad. Res. (IJSAR), eISSN 2583-0279*, vol. 3, no. 10, pp. 1–11, 2023, doi: 10.54756/IJSAR.2023.V3.10.1.

[10]    V. Monjezi, A. Trivedi, G. Tan, and S. Tizpaz-Niari, "Information-theoretic testing and debugging of fairness defects in deep neural networks," in *2023 IEEE/ACM 45th International Conference on Software Engineering (ICSE)*, IEEE, 2023, pp. 1571–1582. doi: 10.1109/ICSE48619.2023.00136.

[11]    M. Salehi, N. Javadpour, B. Beisner, M. Sanaei, and S. B. Gilbert, "Innovative Cybersickness Detection: Exploring Head Movement Patterns in Virtual Reality," *arXiv Prepr. arXiv2402.02725*, 2024, doi: 10.48550/arXiv.2402.02725.

[12]    M. Sanaei, S. B. Gilbert, N. Javadpour, H. Sabouni, M. C. Dorneich, and J. W. Kelly, "The Correlations of Scene Complexity, Workload, Presence, and Cybersickness in a Task-Based VR Game," *arXiv Prepr. arXiv2403.19019*, 2024, doi: 10.48550/arXiv.2403.19019.

[13]    H. Ajami, M. K. Nigjeh, and S. E. Umbaugh, "Unsupervised white matter lesion

identification in multiple sclerosis ( MS ) using MRI segmentation and pattern classification : a novel approach with CVIPtools," vol. 12674, pp. 1–6, 2023, doi: 10.1117/12.2688268.

[14] M. Bodaghi, M. Hosseini, and R. Gottumukkala, "A Multimodal Intermediate Fusion Network with Manifold Learning for Stress Detection," *arXiv Prepr. arXiv2403.08077*, 2024, doi: 10.48550/arXiv.2403.08077.

[15] H. Liu *et al.*, "MEMS piezoelectric resonant microphone array for lung sound classification," *J. Micromechanics Microengineering*, vol. 33, no. 4, p. 44003, 2023.

[16] V. Mohammadi *et al.*, "Development of a Two-Finger Haptic Robotic Hand with Novel Stiffness Detection and Impedance Control," *Sensors*, vol. 24, no. 8, p. 2585, 2024, doi: 10.3390/s24082585.

[17] M. Ebrahimkhani, A. A. Ghiri, and F. Farahmand, "Finite Element Analysis of A Hip Joint Prosthesis with An Extramedullary Fixation System," in *2020 27th National and 5th International Iranian Conference on Biomedical Engineering (ICBME)*, IEEE, 2020, pp. 177–181. doi: 10.1109/ICBME51989.2020.9319421.

[18] M. Farhang and F. Safi-Esfahani, "Recognizing mapreduce straggler tasks in big data infrastructures using artificial neural networks," *J. Grid Comput.*, vol. 18, no. 4, pp. 879–901, 2020, doi: 10.1007/s10723-020-09514-2.

[19] M. K. Nigjeh, H. Ajami, and S. E. Umbaugh, "Automated classification of white matter lesions in multiple sclerosis patients ' MRI images using gray level enhancement and deep learning," vol. 12674, pp. 1–6, 2023, doi: 10.1117/12.2688269.

[20] V. Badrinarayanan, A. Kendall, and R. Cipolla, "Segnet: A deep convolutional encoder-decoder architecture for image segmentation," *IEEE Trans. Pattern Anal. Mach. Intell.*, vol. 39, no. 12, pp. 2481–2495, 2017.

[21] L. Li, T. Zhou, W. Wang, J. Li, and Y. Yang, "Deep hierarchical semantic segmentation," in *Proceedings of the IEEE/CVF Conference on Computer Vision and Pattern Recognition*, 2022, pp. 1246–1257.

[22] B. Bahrami and H. Arbabkhah, "Enhanced Flood Detection Through Precise Water

Segmentation Using Advanced Deep Learning Models," 2024.

[23] S. Eppel, "Hierarchical semantic segmentation using modular convolutional neural networks," *arXiv Prepr. arXiv1710.05126*, 2017.

[24] D. Yang, Y. Du, H. Yao, and L. Bao, "Image semantic segmentation with hierarchical feature fusion based on deep neural network," *Conn. Sci.*, vol. 34, no. 1, pp. 1772–1784, 2022.

[25] M. A. Labbaf-Khaniki, M. Manthouri, and H. Ajami, "Twin Transformer using Gated Dynamic Learnable Attention mechanism for Fault Detection and Diagnosis in the Tennessee Eastman Process," *arXiv Prepr. arXiv2403.10842*, 2024, doi: 10.48550/arXiv.2403.10842.

[26] M. A. L. Khaniki, M. Mirzaeibonehkhater, and M. Manthouri, "Enhancing Fault Detection in Induction Motors using LSTM-Attention Neural Networks," in *2023 9th International Conference on Control, Instrumentation and Automation (ICCIA)*, 2023, pp. 1–5. doi: 10.1109/ICCIA61416.2023.10506369.

[27] Q. Zhao, J. Liu, Y. Li, and H. Zhang, "Semantic segmentation with attention mechanism for remote sensing images," *IEEE Trans. Geosci. Remote Sens.*, vol. 60, pp. 1–13, 2021.

[28] R. Ranjbarzadeh, A. Bagherian Kasgari, S. Jafarzadeh Ghoushchi, S. Anari, M. Naseri, and M. Bendechache, "Brain tumor segmentation based on deep learning and an attention mechanism using MRI multi-modalities brain images," *Sci. Rep.*, vol. 11, no. 1, pp. 1–17, 2021.

[29] J. Kang, L. Liu, F. Zhang, C. Shen, N. Wang, and L. Shao, "Semantic segmentation model of cotton roots in-situ image based on attention mechanism," *Comput. Electron. Agric.*, vol. 189, p. 106370, 2021.

[30] G. Gaál, B. Maga, and A. Lukács, "Attention U-net based adversarial architectures for chest X-ray lung segmentation," *CEUR Workshop Proc.*, vol. 2692, pp. 1–7, 2020, doi: 10.48550/arXiv.2003.10304.

[31] T. Rahman *et al.*, "Reliable tuberculosis detection using chest X-ray with deep learning, segmentation and visualization," *IEEE Access*, vol. 8, pp. 191586–191601, 2020, doi:

10.1109/ACCESS.2020.3031384.

[32] M. A. L. Khaniki and M. Manthouri, "A Novel Approach to Chest X-ray Lung Segmentation Using U-net and Modified Convolutional Block Attention Module," *arXiv Prepr. arXiv2404.14322*, 2024.